\documentclass{article}
\usepackage{amsmath}
\usepackage{amssymb}
\usepackage{amsfonts}

\begin{document}

\title{Spinors and the quaternionic Poincar\'{e} group}
\author{R. Vilela Mendes\thanks{%
rvilela.mendes@gmail.com; rvmendes@ciencias.ulisboa.pt;
https://label2.tecnico.ulisboa.pt/vilela/} \\
CEMS-UL, Faculdade de Ci\^{e}ncias, Univ. Lisboa}
\date{ }
\maketitle

\begin{abstract}
When four dimensional spacetime $\mathcal{R}$ is considered as locally
embedded on a larger manifold $\mathcal{M}$, labelled by higher division
algebra coordinates, a natural question to ask is how much of the symmetry
properties of the larger space are inherited by $\mathcal{R}$. Here this
question is studied when $\mathcal{M}$ is a quaternion manifold. Of
particular relevance is the absence of spinors in the linear representations
of the symmetry group of the larger manifold and the emergence of new
quantum numbers when, by Whitney sums, spinors are implemented on the vector
bundles associated to the coset manifolds of the symmetry groups of $%
\mathcal{M}$. A possible relation to the structures of the standard model is
briefly discussed.     
\end{abstract}

\section{Introduction}

When four dimensional spacetime $\mathcal{R}$ is considered as one real $4-$%
plane in the Grassmanian of a manifold $\mathcal{M}_{\mathbb{K}}$, with
coordinates labelled by a higher division algebra $\mathbb{K}$, a natural
question to ask is how much of the symmetry properties of $\mathcal{M}_{%
\mathbb{K}}$ are inherited by $\mathcal{R}$. The local symmetry group of $%
\mathcal{R}$ is the real Poincar\'{e} group $\mathcal{P}_{\mathbb{R}}$,
whereas for for $\mathcal{M}_{\mathbb{K}}$ it is natural to assume, as local
symmetry group $\mathcal{P}_{\mathbb{K}}$, a group that reduces to $\mathcal{%
P}_{\mathbb{R}}$ in each real $4-$plane. That is, it would be the semidirect
product of a $\mathbb{K-}$Lorentz group $\mathcal{L}_{\mathbb{K}}$ 
\begin{equation*}
\Lambda ^{\dag }g\Lambda =g\hspace{1cm}\Lambda \in \mathcal{L}_{\mathbb{K}}
\end{equation*}%
with metric $g=\left( 1,-1,-1,-1\right) $ together with the translations in $%
\mathbb{K}$. $\mathcal{L}_{\mathbb{K}}$ is the group that has been called by
several authors \cite{Barut} \cite{Ungar} the Lorentz group with real metric.

A first step in the comparison of the consequences of symmetry under $%
\mathcal{P}_{\mathbb{R}}$ and $\mathcal{P}_{\mathbb{K}}$ with $\mathbb{K=C},%
\mathbb{H},\mathbb{O}$ (complex, quaternion and octonion) was taken in Ref.%
\cite{Vilela-IJMPA} by studying the linear irreducible representations of
the groups. An important difference is found in that half-integer spins are
only present for $\mathcal{P}_{\mathbb{R}}$, not for the complex,
quaternionic or octonionic\footnote{%
Notice that for the octonionic case $\mathcal{P}_{\mathbb{O}}$, because of
non-associativity, the group transformations should be interpreted in the
quasi-algebra \cite{Albu1} sense.} groups. Also, for integer spins, when the
discrete transformations are included, there is a superselection rule, but
that will not be of concern in this paper.

At this stage, two points of view are possible:

1. The real $4-$planes, Lorentzian fibers in the Grassmannian $Gr_{4}\left( 
\mathcal{M}_{\mathbb{K}}\right) $ of $\mathcal{M}_{\mathbb{K}}$, do not
inherit the symmetries of the higher algebra, or

2. The symmetries associated to the higher algebra are implemented in a
nonlinear way.

The second hypothesis may be realized in the following way:

The isotropy group $G_{\mathbb{K}}$ of the eigenstates of the first Casimir
operator of $\mathcal{P}_{\mathbb{K}}$, $P_{\mathbb{K}}^{2}$ (Appendix A),
always contain, as a subgroup, the isotropy group $G_{\mathbb{R}}\circeq H$
of the eigenstates of first Casimir operator of $\mathcal{P}_{\mathbb{R}}$.
Using a coset decomposition $G_{\mathbb{K}}/H$, the total space of the group 
$G_{\mathbb{K}}$ may be interpreted as a principal bundle $\left( G_{\mathbb{%
K}},H,G_{\mathbb{K}}/H\right) $\footnote{%
Here the following notation will always be used for bundles (Total space,
Structure Group, Base)$\ $} with structure group $H$ and base $G_{\mathbb{K}%
}/H$. Then, compatibility of the symmetry structures in $\mathcal{M}_{%
\mathbb{K}}$ and in the real Lorentzian fiber $\mathcal{R}$ requires:

A. The implementation of a spin structure over the manifold $G_{\mathbb{K}%
}/H $

B. The matching, by composition, of the representations of $G_{\mathbb{K}}$
and $H$.

The base $B$ of a bundle is a spin manifold if and only if its second
Stiefel-Whitney $w_{2}\left( B\right) $ class vanishes. Notice however, that
the existence of a trivial or nontrivial $w_{2}\left( B\right) $ class only
says whether the tangent bundle $\mathcal{T}B$ over the manifold may or may
not be lifted to a spin manifold. The Stiefel-Whitney classes are
characterizations of the bundles, not of the base manifolds\footnote{%
For example $H^{1}\left( S^{1}\right) \neq 0$, $w_{1}\neq 0$ for the M\"{o}%
bius strip, but $w_{1}=0$ for the cylinder bundle.}. Of course, if the
second cohomology $H^{2}\left( B\right) $ vanishes, then also $0=w_{2}\left(
B\right) \in H^{2}\left( B\right) $. In any case, when the second cohomology
of $B$ is not zero and $w_{2}\left( B\right) \neq 0$, a good indication that
bundles, other than the tangent bundle, are not spin manifolds is the
non-existence of spinors in the linear representations of their total spaces.

In any case the topological obstructions arising from the second cohomology
of $G_{\mathbb{K}}/H$ (condition A) are only indicative of the need for
carrying out Whitney sums to implement spinor states, they do not say, for
example, what the multiplicity of these sums must be. It is condition B that
provides all the required information on the consequences of the
implementation of the symmetries of the larger space $\mathcal{M}_{\mathbb{K}%
}$ on the real spacetrime.

This program was fully carried out in Ref.\cite{Vilela-NPB} for the case $%
\mathbb{K}=\mathbb{C}$. In this case, for massive states in $\mathcal{M}_{%
\mathbb{C}}$, the isotropy group $G_{\mathbb{C}}$ is $U\left( 3,\mathbb{C}%
\right) $ whereas $H$ is the subgroup $SO\left( 3\right) $. Therefore the
manifold where spin states must be expressed is $\mathcal{W=}U\left( 3,%
\mathbb{C}\right) /SO\left( 3\right) $. Up to an irrelevant $U\left(
1\right) $ factor this is the well-known Wu manifold \cite{Chen-Th} \cite%
{spinh} which is not a spin manifold, meaning that its tangent bundle $%
\left( T\mathcal{W}\text{,}SO\left( 6\right) ,\mathcal{W}\right) $ structure
cannot be lifted to $\left( T\mathcal{W}\text{,}Spin\left( 6\right) ,%
\mathcal{W}\right) $, the same applying to the reduced bundle $\left( G_{%
\mathbb{C}},SO\left( 3\right) ,\mathcal{W}\right) $. Not being a spin
manifold, nor a $spin^{c}$ manifold, the Wu manifold is however a $spin^{h}$
manifold, meaning that to obtain a spin structure the manifold must be
coupled by a Whitney sum\footnote{%
The Whitney sum of two bundles B$_{1}$ and B$_{2}$ is a bundle $B_{1}\oplus
B_{2}$ over a common base with fibers that are the direct sum of the fibers
in $B_{1}$ and $B_{2}$} to another manifold with structure group $SU\left(
2\right) \simeq Sp\left( 1\right) $. It so happens that, for the $\left(
U\left( 3\right) ,SO\left( 3\right) ,\mathcal{W=}U\left( 3\right) /SO\left(
3\right) \right) $ bundle, the additional $SU\left( 2\right) $ coupled with
the spin representation over $\mathcal{W}$ deliver the spin $1$ needed to
match the representation of $G_{\mathbb{C}}$, together with an extra scalar
state. In this way both condition A and B are satisfied and, as a result,
not only one obtains a matching of the structures of $\mathcal{P}_{\mathbb{C}%
}$ and $\mathcal{P}_{\mathbb{R}}$ but also some extra $SU\left( 2\right) $
quantum numbers are forced on the spinors. As discussed in \cite{Vilela-NPB}
there is also a "color-like" degeneracy associated to the base manifold.

For zero mass states (of the $P_{\mathbb{C}}^{2}$ operator - Appendix A) the
base manifold is $U\left( 2\right) /SO\left( 2\right) $ and, in this case,
the matching of representations also forces the implementation a
Whitney-like sum with new quantum numbers being assigned to the spinor
states of the real spacetime $\mathcal{R}$. There is however no color
degeneracy in the massless case.

The most interesting point about this construction is the mechanism by which
the matching of the kinematical symmetries between a Grassmanian fiber (the
real space $\mathcal{R}$) and the larger space $\mathcal{M}_{\mathbb{C}}$,
imposes internal-like quantum numbers in the fiber. Also, this structure is
evocative of the one-generation standard model, although different. For
example, color states in the standard model would be linear elements in a $%
SU\left( 3\right) $ representation, whereas here they would be sections on a
curved manifold.

As a natural extension to the work in \cite{Vilela-NPB} , this paper
discusses the symmetries that might be inherited from $\mathcal{M}_{\mathbb{H%
}}$, a quaternionic manifold with real metric.

\section{The quaternionic Poincar\'{e} group and Sp(3)/SO(3)}

The Lorentz invariance group in quaternionic spacetime, denoted $\mathcal{L}%
_{\mathbb{H}}$, is 
\begin{equation}
\Lambda ^{\dagger }G\Lambda =G,  \label{2.1}
\end{equation}%
$G$ being the metric $(1,-1,-1,-1)$. This is the $U\left( 1,3,\mathbb{H}%
\right) $ group over the quaternions $\mathbb{H}$. Together with the
inhomogeneous quaternionic translations one obtains the semi-direct product 
\begin{equation}
T_{4\mathbb{H}}\circledS U(1,3,\mathbb{H}),  \label{2.2}
\end{equation}%
the Poincar\'{e} group (with real metric) $\mathcal{P}_{\mathbb{H}}$ in $%
\mathbb{H}^{4}$. Looked at as a complex group, it is a $52-$parameter group.
The generators of its Lie algebra are $\left\{ M_{\mu \nu },N_{\mu \nu
}^{\alpha },K_{\mu },H_{\mu }^{\alpha }\right\} $ $\left( \alpha
=i,j,k\right) $ (Appendix A).

The isotropy group of massive states (eigenstates of $P_{\mathbb{H}}^{2}$)
is $U(3,\mathbb{H})\simeq Sp\left( 3\right) $. The linear irreducible
representations of this group contain no half-integer spin representations
of the subgroup generated by $\left\{ R_{1},R_{2},R_{3}\right\} $ (the
rotation group - Appendix A). Therefore, as in the complex case $T_{4\mathbb{%
C}}\circledS U(1,3,\mathbb{C})$ studied in \cite{Vilela-NPB}, to obtain
compatibility of the physical representations in $\mathcal{M}_{\mathbb{H}}$
and $\mathcal{R}$, consider the coset bundle%
\begin{equation}
\mathcal{B}_{\mathbb{H}}=\left( Sp\left( 3\right) ,O\left( 3\right)
,Sp\left( 3\right) /SO\left( 3\right) \right)  \label{2.3}
\end{equation}%
and a spin representation over the base $Sp\left( 3\right) /O\left( 3\right) 
$, in fact a vector bundle associated to the principal bundle%
\begin{equation}
\mathcal{B}_{\mathbb{H}s}=\left( Sp\left( 3\right) ,Spin\left( 3\right)
,Sp\left( 3\right) /SO\left( 3\right) \right)  \label{2.4}
\end{equation}%
However, to obtain the matching of the half-integer spin representation in $%
\mathcal{B}_{\mathbb{H}s}$ with the spin one representations of $Sp\left(
3\right) $ one should couple to $\mathcal{B}_{\mathbb{H}s}$ , by a Whitney
sum, other $Sp\left( 1\right) \simeq SU\left( 2\right) $ bundles. In the
case of the lowest dimensional representation of $Sp\left( 3\right) $\ one
has two spin one spaces and therefore the Whitney sum should be%
\begin{equation}
\mathcal{B}_{\mathbb{H}s}\oplus \mathcal{B}_{SU\left( 2\right) }\oplus 
\mathcal{B}_{SU\left( 2\right) }^{\prime }  \label{2.5}
\end{equation}%
bringing to the spinor states additional $SU\left( 2\right) $ quantum
numbers and two scalar states. One sees that in the emergence of new quantum
numbers from the higher division algebra, the central role (in particular
the double Whitney sum) is played by the matching of the representation
spaces (condition B). Topological obstructions to spin lifting over the base
manifolds, arising from the nontrivial cohomology of $Sp\left( 3\right)
/SO\left( 3\right) $, also play a role. In the process, the spin states in $%
\mathcal{B}_{\mathbb{H}s}$ acquire additional $SU\left( 2\right) $ quantum
numbers and two scalar states are also generated.

Another feature of this nonlinear implementation of the quaternionic Poincar%
\'{e} symmetry, $\mathcal{P}_{\mathbb{H}}$, is the degeneracy of state
labels in the base of the $Sp\left( 3\right) /SO\left( 3\right) $ manifold.
In the complex case $\mathcal{B}_{\mathbb{C}}=\left( U(3,\mathbb{C}%
),SO\left( 3\right) ,U(3,\mathbb{C})/SO\left( 3\right) \right) $, the $6-$%
dimensional degrees of freedom in the base manifold $U(3,\mathbb{C}%
)/SO\left( 3\right) $ might be evocative of a non-linear version of a
one-generation color-anticolor quark space. Here the dimension of the base
manifold $Sp\left( 3\right) /SO\left( 3\right) $ is $18$. Could it be
associated to a nonlinear version of $3$ generations of color-anticolor
quark states? Interpreted as a coset decomposition of a representation of $%
Sp\left( 3\right) $, the base manifold is a "mass" degeneracy manifold.
However the states associated to the degrees of freedom of $Sp\left(
3\right) /SO\left( 3\right) $ are degenerated as eigenstates of $P_{\mathbb{H%
}}^{2}=K_{\mu }K^{\mu }+\sum_{\alpha =i,j,k}H_{\mu }^{a}H^{a\mu }$ (Appendix
A), not as mass states of $P_{\mathbb{R}}^{2}=K_{\mu }K^{\mu }$ in the real
Poincar\'{e} group $\mathcal{P}_{\mathbb{R}}$. They might have quite
different $P_{\mathbb{R}}$ masses.

\section{"Massless" states and Sp(2)/SO(2)}

For the massless case ($P_{\mathbb{H}}^{2}=0$) , bringing the eigenstates to
the $\left( p^{0},0,0,p^{0}\right) $ form, the isotropy group is generated
by the following $21$ generators (Appendix A)%
\begin{equation*}
\left\{ R_{3};U_{3}^{\alpha };C_{1}^{\alpha };C_{2}^{\alpha
};L_{1}+R_{2};L_{2}-R_{1};M_{1}^{\alpha }+U_{2}^{\alpha };M_{2}^{\alpha
}+U_{1}^{\alpha };M_{3}^{\alpha }+C_{3}^{\alpha }-C_{0}^{\alpha }\right\}
\end{equation*}%
This algebra is a semidirect sum%
\begin{equation*}
\mathcal{L}G_{2}^{q}=N^{q}\Diamond H^{q}
\end{equation*}%
$N^{q}=\left\{ l_{1},l_{2},m_{1}^{\alpha },m_{2}^{\alpha },m_{3}^{\alpha
}\right\} $ and $H^{q}=\left\{ R_{3},U_{3}^{\alpha },C_{1}^{\alpha
}-C_{2}^{\alpha },C_{1}^{\alpha }+C_{2}^{\alpha }\right\} $ $\left( \alpha
=i,j,k\right) $.

The generators $m_{3}^{\alpha }$, commuting with all the other generators,
are constant in an irreducible representation and denoted $\frac{i}{2}\mu
_{i},\frac{i}{2}\mu _{j},\frac{i}{2}\mu _{k}$. The invariant algebra $N^{q}$
consists of a set of overlapping Heisenberg algebras which, on the space of
differentiable functions of $6$ variables $\overrightarrow{\eta }=\eta
_{i},\eta _{j},\eta _{k},\xi _{i},\xi _{j},\xi _{k}$ have the representation 
\cite{Vilela-IJMPA}%
\begin{eqnarray*}
l_{1}\psi \left( \overrightarrow{\eta },\overrightarrow{\xi }\right)
&=&i\left( \eta _{i}+\eta _{j}+\eta _{k}\right) \psi \left( \overrightarrow{%
\eta },\overrightarrow{\xi }\right) \\
m_{1}^{i}\psi \left( \overrightarrow{\eta },\overrightarrow{\xi }\right)
&=&i\left( -\mu _{i}\frac{\partial }{\partial \eta _{i}}+i\frac{\mu _{k}}{%
\mu _{j}}\eta _{j}-i\frac{\mu _{j}}{\mu _{k}}\eta _{k}\right) \psi \left( 
\overrightarrow{\eta },\overrightarrow{\xi }\right) \\
m_{1}^{j}\psi \left( \overrightarrow{\eta },\overrightarrow{\xi }\right)
&=&i\left( -\mu _{j}\frac{\partial }{\partial \eta _{j}}+i\frac{\mu _{i}}{%
\mu _{k}}\eta _{k}-i\frac{\mu _{k}}{\mu _{i}}\eta _{i}\right) \psi \left( 
\overrightarrow{\eta },\overrightarrow{\xi }\right) \\
m_{1}^{k}\psi \left( \overrightarrow{\eta },\overrightarrow{\xi }\right)
&=&i\left( -\mu _{k}\frac{\partial }{\partial \eta _{k}}+i\frac{\mu _{j}}{%
\mu _{i}}\eta _{i}-i\frac{\mu _{i}}{\mu j}\eta _{j}\right) \psi \left( 
\overrightarrow{\eta },\overrightarrow{\xi }\right) \\
l_{2},m_{2}^{\alpha } &\rightarrow &\left( \eta \rightarrow \xi \right) \\
m_{3}^{\alpha }\psi \left( \overrightarrow{\eta },\overrightarrow{\xi }%
\right) &=&\frac{i}{2}\mu _{\alpha }\psi \left( \overrightarrow{\eta },%
\overrightarrow{\xi }\right)
\end{eqnarray*}%
As in the complex case there are two classes of representations

I - For a nontrivial representation of $N^{q}$, as above, the states are
labelled by the functions $\psi \left( \overrightarrow{\eta },%
\overrightarrow{\xi }\right) $. It is the continuous spin case.

II - For a trivial representation of $N^{q}$ the algebra of the isotropy
group is simply $H^{q}$. $H^{q}$ is the algebra of $Sp\left( 2\right) \sim
SO\left( 5\right) $. A Cartan subalgebra is $\left\{
iR_{3},iC_{+}^{i}=i\left( C_{1}^{i}+C_{2}^{i}\right) \right\} $ and the root
vectors are 
\begin{eqnarray*}
&&iU_{3}^{j}-\varepsilon _{2}U_{3}^{k}-\varepsilon _{1}\left(
C_{-}^{j}+\varepsilon _{2}iC_{-}^{k}\right)  \\
&&iC_{+}^{j}-\varepsilon C_{+}^{k} \\
&&iU_{3}^{i}-\varepsilon C_{-}^{i}
\end{eqnarray*}%
$\varepsilon _{1}$ and $\varepsilon _{2}$ are independent $\pm $ signs. The
first $4$ root vectors have weights $2\left( 
\begin{array}{c}
\varepsilon _{1} \\ 
\varepsilon _{2}%
\end{array}%
\right) $ and the other $4$ have weights $2\left( 
\begin{array}{c}
0 \\ 
\varepsilon 
\end{array}%
\right) $ and $2\left( 
\begin{array}{c}
\varepsilon  \\ 
0%
\end{array}%
\right) $.

In the "massless" complex case $\mathcal{M}_{\mathbb{C}}$ \cite{Vilela-NPB}
one had a $\mathcal{B}_{\mathbb{C}}^{0}=\left( U\left( 2\right) ,SO\left(
2\right) ,U\left( 2\right) /SO\left( 2\right) \right) $ bundle, the $%
SO\left( 2\right) $ fiber being associated to the $R_{3}$ generator, the
only one that is related to the rotation group in the real spacetime $%
\mathcal{R}$. In all the integer or half-integer representations of the $%
H^{c}$ algebra\footnote{$H^{c}=\left\{
R_{3},U_{3}^{i},C_{1}^{i}-C_{2}^{i},C_{1}^{i}+C_{2}^{i}\right\} $}, the
generator $R_{3}$ has integer quantum numbers. Hence, as in the massive
case, to the $\mathcal{B}_{\mathbb{C}}^{0}$ bundle an additional $SU\left(
2\right) $ must be added to obtain both spin representations and the
matching with the full $U\left( 2\right) $ representations.

Here, in the quaternionic case, one obtains a similar two and three-fold
situation. The lowest dimension representation of $SO\left( 5\right) $ has 4
dimensions, with $R_{3}$ quantum numbers $\pm 1$. The coset bundle is%
\begin{equation*}
\mathcal{B}_{\mathbb{H}}^{0}=\left( SO\left( 5\right) ,SO\left( 2\right)
,SO\left( 5\right) /SO\left( 2\right) \right)
\end{equation*}%
with a $9$ dimensional base $SO\left( 5\right) /SO\left( 2\right) $,
Compared with the complex case $\mathcal{B}_{\mathbb{C}}^{0}$ one has a
three-fold replication in the base, suggestive of three generations, and a
two-fold replication in the lowest dimensional representations of the full
spaces ($U\left( 2\right) $ and $SO\left( 5\right) $), requiring, as before,
two additional $SU\left( 2\right) $ bundles in the Whitney sum, as in (\ref%
{2.5}). And no hint of a color degeneracy in the massless case.

\section{Remarks}

a) There have been in the past many attempts to relate the representations
of higher division algebras (\cite{Gunaydin1} \cite{Gunaydin2} \cite%
{Gunaydin4} \cite{Gursey} \cite{Okubo} \cite{Dixon1} \cite{Dixon2} \cite%
{Furey1} \cite{Furey2} \cite{Kugo} \cite{Sudbery} \cite{Evans} \cite%
{Baez-GUT}, etc.) or simply the Clifford algebras \cite{Rowlands} to the
particle states of the standard model. Here, the point of view has been to
consider real spacetime embedded in a higher dimensional manifold, labelled
by division algebra coordinates, and to explore the possible nonlinear
inheritance by the real subspace of the symmetries of the higher dimension
manifold. Whether this embedding is a pure mathematical exercise or it
corresponds to an actual physical phenomenon is an open question. Both
instances are possible because, in the local symmetry groups of the higher
dimension manifolds, spinor states cannot be "rotated" between the real $4-$%
planes and the integer spin states that can be so "rotated", only have
T-violation interactions with the states of the real subspaces, because of a
superselection rule arising from the implementation of the discrete
symmetries \cite{Vilela-IJMPA} \cite{dark}.

b) In any case, mathematical or physical, the embedding in higher dimension
division algebra manifolds is an example of how the so-called internal
quantum numbers may arise from "kinematical" symmetries of a higher
dimension. The quaternionic embedding provides a geometric reason for the
existence of three generations and it is also interesting that scalar field
appear geometrically when the spins over the base manifold are coupled to
the spins of the additional bundles in the Whitney sum.

c) When the nonlinear structures discussed in this paper are, tentatively,
made to correspond to features of the standard model, the color degrees of
freedom are associated to the base in the coset manifold and the flavor
degrees of freedom associated to the additional $SU\left( 2\right) $
manifolds in the Whitney sums. Then, naturally the corresponding gauge
fields would be associated to connections in these manifolds. Notice that in
the quaternionic embedding, in each case, all the generations appear
associated to the same manifold. This might suggest the existence of gauge
fields connecting different generations.

c) All the geometric interpretation done in this paper leaves open the
important dynamical question of the masses of the states, which most
probably depends on the measures on the manifolds.

\section{Appendix A: The quaternionic Poincar\'{e} group}

In the quaternionic Lorentz group, $\mathcal{L}_{\mathbb{H}}\simeq U\left(
1,3,\mathbb{H}\right) $, a representative set of $4\times 4$ matrices,
useful for calculations, related to usual generators by:%
\begin{equation*}
R_{i}=\frac{1}{2}\epsilon _{ijk}M_{jk};\;L_{i}=M_{0i};\;M_{i}^{\alpha
}=N_{0i}^{\alpha };\;U_{i}^{\alpha }=N_{jk}^{\alpha };\;C_{\mu }^{\alpha }=-%
\frac{1}{2}N_{\mu \mu }^{\alpha }g_{\mu \mu }
\end{equation*}%
is%
\begin{equation*}
R_{1}=\left( 
\begin{array}{llll}
0 & 0 & 0 & 0 \\ 
0 & 0 & 0 & 0 \\ 
0 & 0 & 0 & 1 \\ 
0 & 0 & -1 & 0%
\end{array}%
\right) \;R_{2}=\left( 
\begin{array}{llll}
0 & 0 & 0 & 0 \\ 
0 & 0 & 0 & -1 \\ 
0 & 0 & 0 & 0 \\ 
0 & 1 & 0 & 0%
\end{array}%
\right) \;R_{3}=\left( 
\begin{array}{llll}
0 & 0 & 0 & 0 \\ 
0 & 0 & 1 & 0 \\ 
0 & -1 & 0 & 0 \\ 
0 & 0 & 0 & 0%
\end{array}%
\right)
\end{equation*}%
\begin{equation*}
L_{1}=\left( 
\begin{array}{llll}
0 & 1 & 0 & 0 \\ 
1 & 0 & 0 & 0 \\ 
0 & 0 & 0 & 0 \\ 
0 & 0 & 0 & 0%
\end{array}%
\right) \;L_{2}=\left( 
\begin{array}{llll}
0 & 0 & 1 & 0 \\ 
0 & 0 & 0 & 0 \\ 
1 & 0 & 0 & 0 \\ 
0 & 0 & 0 & 0%
\end{array}%
\right) \;L_{3}=\left( 
\begin{array}{llll}
0 & 0 & 0 & 1 \\ 
0 & 0 & 0 & 0 \\ 
0 & 0 & 0 & 0 \\ 
1 & 0 & 0 & 0%
\end{array}%
\right)
\end{equation*}%
\begin{equation*}
U_{1}^{\alpha }=\left( 
\begin{array}{llll}
0 & 0 & 0 & 0 \\ 
0 & 0 & 0 & 0 \\ 
0 & 0 & 0 & \alpha \\ 
0 & 0 & \alpha & 0%
\end{array}%
\right) \;U_{2}^{\alpha }=\left( 
\begin{array}{llll}
0 & 0 & 0 & 0 \\ 
0 & 0 & 0 & \alpha \\ 
0 & 0 & 0 & 0 \\ 
0 & \alpha & 0 & 0%
\end{array}%
\right) \;U_{3}^{\alpha }=\left( 
\begin{array}{llll}
0 & 0 & 0 & 0 \\ 
0 & 0 & \alpha & 0 \\ 
0 & \alpha & 0 & 0 \\ 
0 & 0 & 0 & 0%
\end{array}%
\right)
\end{equation*}%
\begin{equation*}
M_{1}^{\alpha }=\left( 
\begin{array}{llll}
0 & \alpha & 0 & 0 \\ 
-\alpha & 0 & 0 & 0 \\ 
0 & 0 & 0 & 0 \\ 
0 & 0 & 0 & 0%
\end{array}%
\right) \;M_{2}^{\alpha }=\left( 
\begin{array}{llll}
0 & 0 & \alpha & 0 \\ 
0 & 0 & 0 & 0 \\ 
-\alpha & 0 & 0 & 0 \\ 
0 & 0 & 0 & 0%
\end{array}%
\right) \;M_{3}^{\alpha }=\left( 
\begin{array}{llll}
0 & 0 & 0 & \alpha \\ 
0 & 0 & 0 & 0 \\ 
0 & 0 & 0 & 0 \\ 
-\alpha & 0 & 0 & 0%
\end{array}%
\right)
\end{equation*}%
\begin{equation*}
C_{1}^{\alpha }=\left( 
\begin{array}{llll}
0 & 0 & 0 & 0 \\ 
0 & \alpha & 0 & 0 \\ 
0 & 0 & 0 & 0 \\ 
0 & 0 & 0 & 0%
\end{array}%
\right) \;C_{2}^{\alpha }=\left( 
\begin{array}{llll}
0 & 0 & 0 & 0 \\ 
0 & 0 & 0 & 0 \\ 
0 & 0 & \alpha & 0 \\ 
0 & 0 & 0 & 0%
\end{array}%
\right) \;C_{3}^{\alpha }=\left( 
\begin{array}{llll}
0 & 0 & 0 & 0 \\ 
0 & 0 & 0 & 0 \\ 
0 & 0 & 0 & 0 \\ 
0 & 0 & 0 & \alpha%
\end{array}%
\right)
\end{equation*}%
\begin{equation*}
C_{0}^{\alpha }=\left( 
\begin{array}{llll}
\alpha & 0 & 0 & 0 \\ 
0 & 0 & 0 & 0 \\ 
0 & 0 & 0 & 0 \\ 
0 & 0 & 0 & 0%
\end{array}%
\right)
\end{equation*}

$R_{i},L_{i}$ $(i=1,2,3)$ are generators of real rotations and real boosts.
For the other generators:

$\alpha =i,j,k$ with $i^{2}=j^{2}=k^{2}=-1$, $ij=-ji$, $ik=-ki$, $jk=-kj$, $%
ij=k$ and cyclic permutations.

The generators of the real and quaternionic translations are $\left\{ K_{\mu
},H_{\mu }^{a}\right\} $ and the quaternionic "mass squared" operator is%
\begin{equation*}
P_{\mathbb{H}}^{2}=K_{\mu }K^{\mu }+\sum_{\alpha =i,j,k}H_{\mu }^{a}H^{a\mu }
\end{equation*}

Also:

\begin{eqnarray*}
l_{1} &=&L_{1}+R_{2};\;l_{2}=L_{2}-R_{1} \\
m_{1}^{\alpha } &=&M_{1}^{\alpha }+U_{2}^{\alpha };\;m_{2}^{\alpha
}=M_{2}^{\alpha }+U_{1}^{\alpha };\;m_{3}^{\alpha }=M_{3}^{\alpha
}+C_{3}^{\alpha }-C_{0}^{\alpha }
\end{eqnarray*}%
For details and the commutation relations of the isotropy groups of massive
and massless states refer to \cite{Vilela-IJMPA}.

\textbf{Acknowledgments}

Partially supported by Funda\c{c}\~{a}o para a Ci\^{e}ncia e a Tecnologia
(FCT), project UID/04561/2025.


\begin{thebibliography}{99}
\bibitem{Barut} A. O. Barut; \textit{Complex Lorentz group with a real
metric: Group structure}, J. Math. Phys. 5 (1964) 1652-1656.

\bibitem{Ungar} A. A. Ungar; \textit{The abstract complex Lorentz
transformation group with real metric. II. The invariance group of the form} 
$\Vert t\Vert ^{2}-\parallel x\parallel ^{2}$, J. Math. Phys. 35 (1994)
1881-1913.

\bibitem{Vilela-IJMPA} R. Vilela Mendes; \textit{Space-times over normed
division algebras, revisited}, Int. J. Modern Phys. A 35 (2020) 2050055.

\bibitem{Albu1} H. Albuquerque and S. Majid; \textit{Quasialgebra structure
of the octonions}, J. of Algebra 220 (1999) 188-224.

\bibitem{Vilela-NPB} R. Vilela Mendes; \textit{Complex hidden symmetries in
real spacetime and their algebraic structures}, Nucl. Phys. B 1018 (2025)
117046.

\bibitem{Chen-Th} X. Chen; \textit{Bundles of irreducible Clifford modules
and the existence of spin structures}, Ph. Thesis,
https://www.math.stonybrook.edu/alumni/2017-Xuan-Chen.pdf

\bibitem{spinh} M. Albanese and A. Milivojevi\'{c}; \textit{Spin}$^{h}$%
\textit{\ and further generalisations of spin}, Journal of Geometry and
Physics 164 (2021) 104174, 184 (2023) 104709.

\bibitem{Gunaydin1} M. G\"{u}naydin and F. G\"{u}rsey; \textit{Quark
structure and the octonions}, J. Math. Phys. (1973) 1651-1667;

\bibitem{Gunaydin2} M. G\"{u}naydin and F. G\"{u}rsey; \textit{An octonionic
representation of the Poincar\'{e} group}, Lett. Nuovo Cimento 6 (1973)
401-406.

\bibitem{Gunaydin4} M. G\"{u}naydin; \textit{Octonionic Hilbert spaces, the
Poincar\'{e} group and SU(3)}, J. Math. Phys. 17 (1976) 1875-1883.

\bibitem{Gursey} F. G\"{u}rsey and C.-H. Tze; \textit{On the role of
division, Jordan and related algebras in particle physics}, World
Scientific, Singapore 1995.

\bibitem{Okubo} S. Okubo; \textit{Introduction to octonions and other
non-associative algebras in physics}, Cambridge Univ. Press, Cambridge 1995.

\bibitem{Dixon1} G. Dixon; \textit{Division Algebras: Octonions,
Quaternions, Complex Numbers and the Algebraic Design of Physics}, Kluwer,
Dordrecht 1994.

\bibitem{Dixon2} G. Dixon; \textit{Division Algebras, Lattices, Windmill
Tilting}, Createspace Independent Publishing Platform, 2011.

\bibitem{Furey1} C. Furey; $SU(3)_{C}\times SU(2)_{L}\times U(1)_{Y}(\times
U(1)_{X})$ \textit{as a symmetry of division algebraic ladder operators},
Eur. Phys. J. C (2018) 78:375.

\bibitem{Furey2} C.Furey; \textit{Three generations, two unbroken gauge
symmetries, and one eight-dimensional algebra}, Physics Letters B785 (2018)
84--89.

\bibitem{Kugo} T. Kugo and P. Towsend; \textit{Supersymmetry and the
division algebras}, Nucl. Phys. B 221 (1983) 357-380.

\bibitem{Sudbery} A. Sudbery; \textit{Division algebras, (pseudo)orthogonal
groups and spinors}, J. Phys. A Math. Gen. 17 (1984) 939-955.

\bibitem{Evans} J. M. Evans; \textit{Supersymmetric Yang-Mills theories and
division algebra}s, Nucl. Phys. B 298 (1988) 92-108.

\bibitem{Baez-GUT} J. Baez and J. Huerta; \textit{The algebra of grand
unified theories}, Bulletin of the AMS 47 (2010) 483-552.

\bibitem{Rowlands} P. Rowlands and S. Rowlands; \textit{Are octonions
necessary for the standard model?}, J. of Physics Conf. Series 1251(2019)
012044.

\bibitem{dark} R. Vilela Mendes; \textit{T-violation and the dark sector},
Modern Physics Letters A 38 (2023) 2350165.
\end{thebibliography}
\end{document}